# MICROCANONICAL ENSEMBLES OF SYSTEMS WITH MECHANICAL CONSTRAINTS

I.H. UMIRZAKOV

Institute of Thermophysics SB RAS

1 Lavrentyev Ave., Novosibirsk, Russia

E-mail: tepliza@academ.org

## ABSTRACT

*We have obtained an exact expression for the phase-space volume corresponding to a microcanonical ensemble of systems under center of mass, total linear and angular momenta conservation constraints, and arbitrary constraints on the coordinates of particles of the system. Methods are suggested to calculate the phase-space volume and density of states from the mean kinetic energy and mean inverse kinetic energy. Methods to control equilibrium in simulations are also presented. We have derived exact formulae for several thermodynamic response functions. It is shown how to obtain the phase-space volume corresponding to other ensembles when one or several of the constraints are removed. It is shown that the phase-space volume of a system at positive values of the energy is a polynomial function of the energy if the potential energy of interaction between particles of the system consists of a hard-core potential and an arbitrary negative potential. We have also shown that the coefficients of the polynomial function can be determined from simulations.*

**Key words**: *finite system, hard-core potential, phase space volume, response function, equilibrium, fluctuation.*

## 1. INTRODUCTION

Numerical simulation methods such as molecular dynamics (MD) and Monte-Carlo (MC) methods are used as tools for studying properties of a physical system. The simulation methods can be used if the number of particles in the system is finite. Clusters as finite systems are physically interesting [1—7]. Therefore, consideration of microcanonical ensembles of finite systems under various constraints is of great interest [8—11]. In the present work we consider an [$E$, $Q$, $n$, **P**, **L**, **R**] ensemble, where $E$, $Q$, $n$, **P**, **L**, and **R** are the same and constant for all systems of the ensemble. Here $E$ is the total energy of a system of $n$ particles with masses $m_a$, ($a = 1,...,n$), **P** and **L** are the linear and angular momenta of the system, respectively, **R** is the vector of coordinates of the center of mass of the system. Here $Q$ denotes any condition on the coordinates of particles, for example, it is well known that a cluster in vacuo may be inherently unstable with respect to evaporation of atoms from the cluster. In order to avoid evaporation the cluster is usually placed in a container in numerical simulations of clusters [2, 3, 5], in this case $Q = V$, where $V$ is equal to the volume of the container. Alternatively, one can introduce a distance cutoff $r_c$ and take appropriate simulation if an atom attempts to evaporate [4], then $Q$ denotes a condition of connectivity of all particles in the cluster (two particles are connected if $r_{ab} \leq r_c$, where $r_{ab}$ is the distance between particles $a$ and $b$).

## 2. RESULTS AND CONCLUSIONS

The phase-space volume corresponding to the system in phase space is given by [12]

$$\Omega = \tilde{n} \int d^{3n}\mathbf{r}\,\delta(\mathbf{R}' - \mathbf{R}) \int d^{3n}\mathbf{p}\,\theta(E - H)\delta(\sum_{a=1}^{n}\mathbf{p}_a - \mathbf{P})\delta(\sum_{a=1}^{n}\mathbf{r}_a \times \mathbf{p}_a - \mathbf{L}) . \qquad (1)$$

Here $\tilde{n} = (2\hbar\pi)^{-3n}$, $\hbar$ is Planck's constant, $\mathbf{r} = (\mathbf{r}_a, a=1,...,n)$, $\mathbf{p} = (\mathbf{p}_a, a=1,...,n)$, $\mathbf{r}_a$ and $\mathbf{p}_a$ are the vectors of coordinates and linear momenta of particles, respectively, $\mathbf{R}' = \sum_{a=1}^{n} m_a \mathbf{r}_a / M$ is the vector of coordinates of the center of mass of the system, $M = \sum_{a=1}^{n} m_a$ is the total mass, $H = \sum_{a=1}^{n} \frac{\mathbf{p}_a^2}{2m_a} + U$ is a Hamiltonian, $U = U(\mathbf{r})$ is equal to the sum of a potential of interaction of particles with each other and a potential of external forces, $\delta(x)$ is a delta-function, $\theta(x)$ is a Heaviside step function. Laplace transformation of the left- and right-hand sides of Eq. (1) and Fourier expansions of the last two $\delta$-functions in Eq. (1) lead to

$$L_\lambda[\Omega] = \frac{c}{(2\pi)^6 \lambda} \int d^{3n}\mathbf{r}\,\delta(\mathbf{R}' - \mathbf{R}) \int d^3\mathbf{b} \int d^3\mathbf{a} \int d^{3n}\mathbf{p}\,\exp[i\mathbf{a}(\sum_{a=1}^{n}\mathbf{p}_a - \mathbf{P}) + i\mathbf{b}(\sum_{a=1}^{n}\mathbf{r}_a \times \mathbf{p}_a - \mathbf{L}) - \lambda H].$$

We sequentially integrate with respect to the momenta, vectors $\mathbf{a}$ and $\mathbf{b}$, take the inverse Laplace transform of the result, and obtain for $n > 2$

$$\Omega = \tilde{n}_1 \int d^{3n}\mathbf{r}\,\delta(\mathbf{R}' - \mathbf{R}) \frac{(E' - U')^{N/2} \theta(E' - U')}{(I_1 I_2 I_3)^{1/2} \Gamma(N/2 + 1)}, \qquad (2)$$

$$U' = U + \frac{(\mathbf{L} - \mathbf{R}' \times \mathbf{P})\hat{\mathbf{I}}^{-1}(\mathbf{L} - \mathbf{R}' \times \mathbf{P})}{2},$$

$$E' = E - \frac{\mathbf{P}^2}{2M},$$

$$c_1 = \frac{\prod_{a=1}^{n} m_a^{3/2}}{(2\pi)^3 M^{3/2}(2\hbar^2\pi)^{3n/2}},$$

where $N = 3n - 6$ is the number of degrees of freedom, and $I_1 I_2 I_3 = \det \hat{\mathbf{I}}$, and $I_\alpha$ ($\alpha = x, y, z$) are the principal momenta of inertia of the system, and $\hat{\mathbf{I}}$ is the tensor of inertia in the center of mass of a laboratory-oriented frame with matrix elements

$$I_{\alpha\alpha} = \sum_{a=1}^{n} m_a(\mathbf{r}_a'^2 - \alpha_a^2), \quad \mathbf{r}_a' = \mathbf{r}_a - \mathbf{R}' = (x_a, y_a, z_a),$$

$$I_{\alpha\beta} = -\sum_{a=1}^{n} m_a \alpha_a \beta_a, \quad \alpha, \beta \, (\neq \alpha) = x, y, z.$$

Note that the phase-space volumes for [$E$, $Q$, $n$, **P**, **L**], [$E$, $Q$, $n$, **P**, **R**], and [$E$, $Q$, $n$, **L**, **R**] ensembles are obtained by integrating Eq. (2) with respect to **R**, **L**, and **P**. One can find the phase-space volumes for [$E$, $Q$, $n$, **P**], [$E$, $Q$, $n$, **R**] and [$E$, $Q$, $n$, **L**] ensembles by integrating Eq. (2) with respect to (**R**, **L**), (**L**, **P**) and (**R**, **P**). The phase-space volume for the [$E$, $Q$, $n$] ensemble is determined by integrating Eq. (2) with respect to (**R**, **L**, **P**). These integrals are obtained analytically.

Using the following equation

$$Z = \int_0^\infty \rho \exp(-E/kT) dE$$

we can easily obtain the partition function for appropriate canonical ensembles (the energy is measured from its value at the minimum). Here

$$\rho = \partial \Omega / \partial E \tag{3}$$

is the density of states.

Knowing the phase-space volume one can easily obtain the entropy $S$, the microcanonical temperature $T$ and the pressure $p$ by [8, 9, 12]

$$S = k \ln \Omega, \tag{4}$$
$$T = (\partial S / \partial E)^{-1} = \Omega / k\rho, \tag{5}$$
$$p = T \frac{\partial S}{\partial V} = \frac{1}{\rho} \frac{\partial \Omega}{\partial V}. \tag{6}$$

Here $k$ is the Boltzmann constant. We find from Eqs. (2), (3)

$$\rho = \tilde{n}_1 \int d^{3n} \mathbf{r} \delta(\mathbf{R}' - \mathbf{R}) \frac{(E' - U')^{N/2 - 1} \theta(E' - U')}{(I_1 I_2 I_3)^{1/2} \Gamma(N/2)}. \tag{7}$$

From Eqs. (2), (5), and (7) one can obtain

$$<K> = \frac{N}{2} kT = \frac{N}{2} \frac{\Omega}{\rho}, \tag{8}$$

$$<K^{-1}> = \frac{2}{N-2} \frac{1}{\rho} \frac{\partial \rho}{\partial E}, \tag{9}$$

$$<K^2> = \frac{N(N+2)}{4\rho} \int_0^E \Omega dE, \tag{10}$$

where $K = E - \mathbf{P}^2/2M - U'$ is the kinetic energy, and

$$<A> = \tilde{n}_1 \int d^{3n}\mathbf{r}\delta(\mathbf{R}'-\mathbf{R})A\frac{(E'-U')^{N/2-1}\theta(E'-U')}{(I_1 I_2 I_3)^{1/2}\Gamma(N/2)\rho}. \qquad (11)$$

From (8) and (10) we obtain an exact formula for the mean relative fluctuations of the kinetic energy,

$$\Delta_K = \frac{<K^2>-<K>^2}{<K>^2} = \frac{N+2}{N\Omega^2}\rho\int_0^E \Omega dE - 1. \qquad (12)$$

Equation (8) gives

$$\frac{\partial \ln \Omega}{\partial E} = \frac{N}{2<K>}. \qquad (13)$$

We integrate (13) and obtain

$$\Omega = \Omega(E_0)\exp\int_{E_0}^E \frac{N}{2<K>}dE, \qquad (14)$$

$$\rho = \rho(E_0)\frac{<K>(E_0)}{<K>(E)}\exp\int_{E_0}^E \frac{N}{2<K>}dE, \qquad (15)$$

where $E_0$ is some value of the energy, and $\rho(E_0) = N\Omega(E_0)/2<K>(E_0)$, $<A>(E)$ denotes the dependence of $<A>$ on energy $E$.

A good deal of effort was devoted to calculate the density of states $\rho$ [13, 14]. The multiple histogram method [15, 16] and the adiabatic switching method [17] were used. We consider the Eqs. (14), (15) as a tool for determining the phase-space volume $\Omega$ and the density of states $\rho$ from a caloric equation of state, that is, a dependence of $<K>$ on the energy $E$ if the latter is known from simulations. This tool was first suggested in [18] and successfully used to determine $\Omega(E)$ and $\rho(E)$ of $LJ_{13}$ cluster in order to estimate the evaporation rate of particles from the cluster. If the mean kinetic energy $<K>(E)$ is determined from simulations, one can calculate the mean relative fluctuations of the kinetic energy, $\Delta_K(E)$, using Eqs. (12), (14), and (15). Then by comparing the calculated $\Delta_K(E)$ with that obtained from simulations one can determine the degree of equilibrium in the system, because the quantities must be equal to each other if conditions of *the hypothesis of equal a priori probabilities* are provided in the simulations. This method was used in [18-20] to control equilibrium in small clusters.

Equation (9) gives

$$\frac{\partial \ln \rho}{\partial E} = \frac{N-2}{2}<K^{-1}>. \qquad (16)$$

We integrate (16) and obtain

$$\rho = \rho(E_0)\exp\int_{E_0}^{E}\frac{N-2}{2}<K^{-1}>dE, \qquad (17)$$

$$\Omega = \rho(E_0)\int_0^E[\exp\int_{E_0}^{E}\frac{N-2}{2}<K^{-1}>dE]dE. \qquad (18)$$

We suggest to use Eqs. (17) and (18) to determine $\Omega(E)$ and $\rho(E)$ from a dependence of the mean of inverse kinetic energy $<K^{-1}>$ on the energy $E$. If $<K^{-1}>(E)$ is determined by a simulation method, one can calculate the mean relative fluctuations of the kinetic energy, $\Delta_K(E)$, by using Eqs. (12), (17) and (18). Then by comparing the calculated $\Delta_K(E)$ with that obtained from simulations one can determine the degree of equilibrium in the system. We note also that if $<K>(E)$ and $<K^{-1}>(E)$ are determined by simulations, one can calculate the density of states using two ways, i.e., using Eq. (15) or (l7). Then by comparing the calculated densities of states with each other one can determine the degree of equilibrium in the system. Equations similar to Eqs. (8)-(18) can be obtained for other microcanonical ensembles and used to determine $\Omega$ and $\rho$, and to control equilibrium in simulations of these ensembles.

It is necessary to determine $\rho(E_0)$ in Eqs. (14), (15) and (17), (18) to obtain the absolute values of the phase-space volume and density of states. Fortunately most of practically interesting systems have global minima on the potential energy surface $U'=U'(\mathbf{r})$, and for such systems $\rho(E_0)$ can be determined explicitly at $E_0 \to 0$. Indeed, the harmonic approximation can be used at a global minimum at small energies. The equations

$$\Omega_h = \frac{E^N}{N!(\hbar\omega)^N}, \qquad \rho_h = \frac{E^{N-1}}{(N-1)!(\hbar\omega)^N},$$

are valid in the harmonic approximation (the energy is measured from its value at the minimum) [3]. Here $\omega^N = \prod_{i=1}^{N}\omega_i$, and $\omega_i$ are the frequencies of harmonic oscillators. Using (13) and (l6) we obtain for a system of $N$ harmonic oscillators

$$<K_h> = \frac{E}{2}, \qquad <K_h^{-1}> = \frac{N-1}{N-2}\frac{2}{E}.$$

We find from Eqs. (13)-(15) and (16)-(18)

$$\Omega = \Omega_h \exp\int_0^E \frac{N}{2}(\frac{1}{<K>}-\frac{1}{<K_h>})dE, \qquad (19)$$

$$\rho = \rho_h \frac{E}{2<K>}\exp\int_0^E \frac{N}{2}(\frac{1}{<K>}-\frac{1}{<K_h>})dE, \qquad (20)$$

$$\Omega = \int_0^E[\rho_h \exp\int_0^E \frac{N-2}{2}(<K^{-1}>-<K_h^{-1}>)dE]dE, \qquad (21)$$

$$\rho = \rho_h \exp\int_0^E \frac{N-2}{2}(<K^{-1}>-<K_h^{-1}>)dE. \qquad (22)$$

If conditions of *the hypothesis of equal a priori probabilities* are provided in MD or MC simulations, i.e., the system is in equilibrium conditions, the following identity, obtained from Eqs. (20) and (22), must be satisfied

$$\int_0^E \frac{N-2}{2}(<K^{-1}> - <K_h^{-1}>)dE = \ln\frac{E}{2<K>} + \int_0^E \frac{N}{2}(\frac{1}{<K>} - \frac{1}{<K_h>})dE. \quad (23)$$

Deviation of a system from Eq. (23) means that the system does not reach equilibrium conditions.

For an ensemble of systems of *n* particles occupying a volume $V$ ($Q = V$), i.e. for an [$E$, $V$, $n$, **P**, **L**, **R**] ensemble it is convenient to scale the coordinates, $\mathbf{r} = \mathbf{r}/V^{1/3}$, in Eq. (2), which transforms $\Omega$ into the form

$$\Omega = \tilde{n}_1 V^{n'} \int d^{3n}\mathbf{r}\, \delta(\mathbf{R}' - \mathbf{R}) A \frac{(E' - U'(V,\mathbf{r}))^{N/2} \theta(E' - U'(V,\mathbf{r}))}{(I_1 I_2 I_3)^{1/2} \Gamma(N/2+1)},$$

where **r** satisfies $U'(V,\mathbf{r}) \le E'$, $n' = n-2$, and

$$U'(V,\mathbf{r}) = U(V^{1/3}\mathbf{r}) + \frac{\mathbf{L}'\hat{\mathbf{I}}^{-1}\mathbf{L}'}{2V^{2/3}}, \quad \mathbf{L}' = \mathbf{L} - V^{1/3}\mathbf{R}' \times \mathbf{P},$$

$\mathbf{R}'$, $\mathbf{R}$, $\bar{\mathbf{I}}$ and $\bar{I}_\alpha$ ($\alpha = 1,2,3$) do not depend on *V*. Equation (6) leads to

$$p = \frac{n'kT}{V} - <\frac{\partial U'}{\partial V}> = \quad (24)$$

$$= \frac{2n'<K>}{NV} - <\frac{\partial U}{\partial V}> + \mathbf{L}\frac{<\hat{\mathbf{I}}^{-1}\mathbf{L}> - V^{1/3}<\hat{\mathbf{I}}^{-1}\mathbf{R}' \times \mathbf{P}>}{3V^{5/3}}. \quad (25)$$

The last term in Eq. (25) for the pressure disappears at **L**=0, and at $<\hat{\mathbf{I}}^{-1}\mathbf{L}> = V^{1/3}<\hat{\mathbf{I}}^{-1}\mathbf{R}' \times \mathbf{P}>$. From Eqs. (5) and (24) we obtain exact formulae for thermodynamic response functions, that is, the heat capacity $\tilde{N}_V$, Gruneisen parameter $\gamma = V\partial p/\partial E$ and bulk modulus $B_S = -V(\partial p/\partial V)_S$

$$\frac{k}{C_V} = \left(\frac{\partial kT}{\partial E}\right)_V = 1 - (1 - 2/N) \cdot <K> \cdot <K^{-1}>,$$

$$\gamma = \frac{n'k}{C_V} + (1 - N/2) \cdot V \cdot [<K^{-1} \cdot \frac{\partial U'}{\partial V}> - <K^{-1}> \cdot <\frac{\partial U'}{\partial V}>],$$

$$B_S = \frac{2n'<K>}{NV} \cdot (1 + 2\gamma - \frac{n'k}{C_V}) + V \cdot <\frac{\partial^2 U'}{\partial V^2}> -$$

$$- (N/2 - 1) \cdot V \cdot [<K^{-1} \cdot \left(\frac{\partial U'}{\partial V}\right)^2> - <K^{-1} \cdot \frac{\partial U'}{\partial V}> \cdot 2 \cdot <\frac{\partial U'}{\partial V}> + \left(<\frac{\partial U'}{\partial V}>\right)^2 \cdot <K^{-1}>].$$

[*E, V, n*] and [*E, V, n*, **P**] ensembles were earlier treated in [8, 9] and [10], respectively. An expression for the density of states corresponding to a [*E, V, n*, **P, L**] ensemble was earlier obtained in [11].

We note that conservation of coordinates of the center of mass of a system can appreciably change the properties of the microcanonical caloric curve and the equation of state of a small finite system (see [21] where exact analytical expressions for the phase-space volume and density of states of a system of two particles interacting by both Lennard-Jones and Morse potentials in a spherical volume were obtained).

Let us consider a special case of potential energy, $U(\mathbf{r}_1,...,\mathbf{r}_n) = +\infty$ if some $r_{ab} \leq \sigma_{ab}$, and $U(\mathbf{r}_1,...,\mathbf{r}_n) \leq 0$ if otherwise ($\sigma_{ab}$ are constants). We integrate Eq. (2) with respect to (**R, L, P**) and obtain for the phase-space volume corresponding to a [*E, Q, n*] ensemble

$$\Omega = \frac{M^{3/2} c_1}{(2\pi)^{-3}} \int d^{3n}\mathbf{r} \frac{(E-U)^{3n/2} \theta(E-U)}{\Gamma(3n/2+1)}.$$

This expression takes the following form at $E \geq 0$

$$\Omega = \frac{M^{3/2} c_1}{(2\pi)^{-3}} \int d^{3n}\mathbf{r} \frac{(E-U)^{3n/2}}{\Gamma(3n/2+1)} \prod_{a=1}^{n-1} \prod_{b=a+1}^{n} \theta(r_{ab} - \sigma_{ab}).$$

For an even number of particles $n$ we expand $(E-U)^{3n/2}$ as a polynomial function of the energy $E$ with coefficients $A_l$ which does not depend on the energy ($L=3n/2$)

$$\Omega = \sum_{l=0}^{L} A_l E^l,$$

$$A_l = \frac{C_L^l M^{3/2} c_1}{(2\pi)^{-3} L!} \int d^{3n}\mathbf{r} (-U)^{L-l} \prod_{a=1}^{n-1} \prod_{b=a+1}^{n} \theta(r_{ab} - \sigma_{ab}).$$

We find from (8)

$$<K>(E) = \frac{\sum_{l=0}^{L} A_l E^l}{\sum_{l=1}^{L} l A_l E^{l-1}} \frac{3n}{2}.$$

If we find from simulations the mean kinetic energy $<K>(E)$ for $L+1$ different values of the energy $E_\alpha$ ($\alpha = 0,...,L$) and the following system of $L+1$ linear equations

$$A_0 + \sum_{l=1}^{L} A_l [E_\alpha^l - l E_\alpha^{l-1} \frac{2 <K>(E_\alpha)}{3n}] = 0, \quad \alpha = 0,...,L,$$

has a single solution, we can determine the $L+1$ coefficients $A_l$ ($l = 0, ..., L$) as the solution of this system of equations. Having known all $A_l$ we can determine the phase-space volume for arbitrary $E \geq 0$. All coefficients $A_l$ can be also determined by interpolating the caloric curve $<K>(E)$ in some interval of energy at $E \geq 0$ by Eq. (27). Equation (26) can be used to determine thermodynamic properties of gases, fluids and solids at high energies.

Expansions similar to (26) can be obtained for [*E, Q, n*, **P, L** = 0, **R**], [*E, Q, n*, **P, L** = 0] and [*E, Q, n*, **R**] ensembles if *n* is even and the coefficients of these expansions can be obtained from simulations in a similar way as for an [*E, Q, n*] ensemble. If the number of particles *n* is odd, we can obtain expansions similar to (26) for [*E, Q, n*, **P, R**], [*E, Q, n*, **L** = 0, **R**], [*E, Q, n*, **L** = 0] and [*E, Q, n*, **P**] ensembles, and the coefficients of these expansions can be obtained from simulations in a similar way as for an [*E, Q, n*] ensemble.

We have obtained an exact expression for the phase-space volume corresponding to a microcanonical ensemble of systems under center of mass, total linear and angular momenta conservation constraints, and arbitrary constraints on the coordinates of particles of the system. Methods are suggested to calculate the phase-space volume and density of states from the mean kinetic energy and mean inverse kinetic energy. Methods to control equilibrium in simulations are also presented. We have derived exact formulae for several thermodynamic response functions. It is shown how to obtain the phase-space volume corresponding to other ensembles when one or several of the constraints are removed. It is shown that the phase-space volume of a system at positive values of the energy is a polynomial function of the energy if the potential energy of interaction between particles of the system consists of a hard-core potential and an arbitrary negative potential. We have also shown that the coefficients of the polynomial function can be determined from simulations.

**REFERENCES**

1. Bixon, M. and Jortner, J., Energetic and thermodynamic size effects in molecular clusters, *J. Chem. Phys.*, **91**, P. 1631—1642, 1989.
2. Tsai, C.J. and Jordan, K.D., Use of histogram and jump-walking methods for overcoming slow barrier crossing behavior in Monte Carlo simulations: application to the phase transitions, *Ibid.*, **99**, P. 6957-6970, 1993.
3. Wales, D.J., Coexistence in small inert gas clusters, *Mol. Phys.*, **78**, P. 151-171, 1993.
4. Chekmarev, S.F. and Umirzakov, I.H., An analytic model for atomic clusters, *Z. Phys. D.*, **26**, P. 373-376, 1993.
5. Lynden-Bell, R.M. and Wales, D.J., Free energy barriers to melting in atomic clusters, *J. Chem. Phys.*, **101**, P. 1460-1476, 1994.
6. Wales, D.J. and Berry, R.S., Coexistence in finite systems, *Phys. Rev. Lett.*, **73**, P. 2875-2878, 1994.
7. Umirzakov, I.H., Van der Waals type loop in microcanonical caloric curves of finite systems, *Phys. Rev. E*, **60**, P. 7550-7553, 1999.
8. Munster, A., Statistical Thermodynamics, Springer-Verlag, Berlin, 1969, Vol. **I**, Chap. 1.
9. Pearson, E.M., Halicioglu, T., and Teller, W.A., Laplace-transform technique for deriving thermodynamic eduations from the classical microcanonical ensemble, *Phys. Rev. A*, **32**, P. 3030-3039, 1985.

10. Cagin, T. and Ray, J.R., Fundamental treatment of molecular-dynamics ensembles, *Ibid*., **37**, P. 247-251, 1988.
11. Nyman, G., Nordholm, S., and Schranz, H.W., Efficient microcanonical sampling for a preselected total angular momentum, *J. Chem. Phys*., **93**, P. 6767-6773, 1990.
12. Landau, L D. and Lifshits, E M., Statistical Physics, Pergamon, London, 1965.
13. Labastie, P. and Whetten, R.L., Statistical thermodynamics of the cluster solid-liquid transition, *Phys. Rev. Lett*., **65**, P. 1567-1570, 1990.
14. Weerasinghe, S. and Amar, F.G., Absolute classical densities of states for very anharmonic systems and applications to the evaporation of rare gas clusters, *J. Chem. Phys*., **98**, P. 4967-4983, 1993.
15. Ferrenberg, A.M. and Swendsen, R.H., New Monte Carlo technique for studying phase transitions, *Phys. Rev. Lett*., **61**, P. 2635-2638, 1988.
16. Iidem N., Optimized Monte Carlo analysis, *Ibid.,* **63**, P. 1195-1198, 1989.
17. Reinhard, W.P. and Hanter, J.E., Variational path optimization and upper and lower bounds to free energy changes via finite time minimization of external work, *J. Chem. Phys*., **97**, P. 1599-1601, 1992.
18. Chekmarev, S.F., Umirzakov, I.H., and Liu, F.S., New Statistical Model for Atomic Clusters Taking into Account Transition States, Preprint, Institute of Thermophysics, Novosibirsk, 1991.
19. Umirzakov, I.H., Finite Systems of Particles Interacting via Van der Waals forces, Ph. D. Thesis, Institute of Thermophysics, Novosibirsk, 1993.
20. Chekmarev, S.F. and Krivov, S.V., Total and fractional densities of states from caloric relations, *Phys. Rev E*, **57**, P. 2445-2448, 1998.
21. Umirzakov, I.H., Exact equilibrium statistical mechanics of two particles interacting via Lennard-Jones and Morse potentials, *Ibid*., **61**, 7188-7191, 2000.